# VARIABLE SECOND-ORDER INCLUSION PROBABILITIES AS A TOOL TO PREDICT THE SAMPLING VARIANCE

Bastiaan Geelhoed; Delft University of Technology, Mekelweg 15, 2629 JB Delft, The Netherlands; b.geelhoed@tudelft.nl

## ABSTRACT

A generalization of Gy's theory for the variance of the fundamental sampling error is reviewed. Practical situations where the generalized model potentially leads to more accurate variance estimates are identified as: clustering of particles, differences in densities or sizes of the particles or repulsive inter-particle forces. Two general approaches for estimating an input parameter for the generalized model are discussed. The first approach consists of modelling based on physical properties of particles such as size, density and electrostatic forces between particles. The second approach uses image analysis of actual samples. Further research into both methods is proposed and a suggestion is made to use line-intercept sampling combined with Markov Chain modelling in the second approach.

It is concluded that although, at the moment, it is too early for a routine application of the generalized theory, the generalization has the potential of providing more accurate variance estimates than are possible in the theory of Gy. Therefore, further research into the development and expansion of the generalized theory is worthwhile.

## INTRODUCTION

During the previous World Conference on Sampling and Blending, WCSB-2, a generalization of Gy's model for the fundamental sampling error (see Gy 1979, 1983) was proposed and an equation to predict the variance of the fundamental sampling error was derived (Geelhoed, 2005A). It was discussed that application of this equation could, at least in theory, lead to more accurate predictions of the variance of the fundamental error than are currently provided in the theory of Gy. In this contribution, the generalization and the new theoretical developments since it was proposed at WCSB-2 will be reviewed. Situations where it is expected that the generalized theory will become relevant will be discussed. An important new input parameter of the new model is the parameter for the dependent selection of particles. Therefore, two general approaches of estimating this input parameter will be discussed. The first approach is a modelling approach based on physical properties of particles such as size, density and electrostatic forces between particles. The second approach uses image analysis of actual samples. Further research into both approaches will be proposed.

## METHODOLOGY

Gy's theory is based on the underlying assumption that sampling corresponds to Poisson sampling: every particle is independently subjected to a Bernoulli experiment, where the i-th particle in the batch has a probability of $q_i$ of becoming part of the sample and a probability of $1-q_i$ of not becoming part of the sample. If sampling correctness is assumed, which requires that all particles of the sampling target have an equal probability of becoming part of the sample, all probabilities $q_i$ become equal and can therefore be denoted by a single symbol q. The parameter q is generally quantified (estimated) using the ratio of the sample size and the batch size from which the sample was taken, which can be interpreted as the first-order inclusion probabilities of the particles.

A generalization of the above model was proposed (Geelhoed, 2005A). The generalization concerns the second-order inclusion probabilities, i.e. the probabilities that a pair of particle i and particle j becomes part of the sample. Because the theory of Gy is based on Poisson sampling, the second-order inclusion probabilities are given in Gy's theory by $q_i \times q_j$. In the generalized model, the second-order inclusion probability, denoted as $\pi_{ij}$, is given by:

$$\pi_{ij} = q_i \times q_j \times (1-C_{ij}) \qquad (Eq.\ 1)$$

where $C_{ij}$ is the "parameter for the dependent selection of particles", which forms a symmetric matrix ($C_{ij} = C_{ji}$). It is noted that the values for $q_i$, $q_j$ and $C_{ij}$ are generally influenced by the choice of the sampling strategy, the sample mass and the particle properties, so the second-order inclusion probabilities (and ultimately the sampling variance) will also depend on these circumstances. An equation for the variance of the fundamental sampling error was derived using (1) and the following five general and practically reasonable conditions (Geelhoed, 2005A):

i.   The particles can be classified using a finite number of classes.
ii.  The size of the batch from which the sample is drawn is infinite, so that sampling can effectively be regarded as "sampling with replacement".
iii. The first-order inclusion probabilities of particles may vary between classes, but do not vary for particles within a class.
iv.  The second-order inclusion probabilities may depend on the classes of both particles, but there is no variation in second-order inclusion probabilities of different particles pairs but with each member belonging to the same class as the corresponding member in the other particle pair.

v.  Variations in sample mass remain small, so that the sample concentration can effectively be linearized as a function of the sample mass.

Under these conditions, the equation for the variance of the fundamental sampling error is equivalent to the following equation for the variance of the concentration in the sample, $V(c_{sample})$:

$$V(c_{sample}) = \frac{1}{M'^2_s} \sum_i N'_i m_i^2 (c_i - c')^2 - \frac{1}{M'^2_s} \sum_i \sum_j C_{ij} N'_i N'_j m_i m_j (c_i - c'_s)(c_j - c'_s)$$

(Eq. 2)

Where the parameters $m_i$ and $c_i$ denote respectively the mass of and concentration in a particle belonging to the i-th particle class, $M'_s$, $N'_i$ and $c'_s$ are respectively the expected value of the sample mass, the expected value of the number of particles belonging to the i-th class in the sample and the expected value of the concentration of the property of interest in the sample. In (2) and subsequent parts of this article, $C_{ij}$ represents the parameter for the dependent selection of particles of a pair consisting of a particle belonging to the i-th class and a particle belonging to the j-th class.

The first term on the right-hand side of (2) corresponds to the equation derived in the theory of Gy (Geelhoed, 2005A), where the variance is inversely proportional to the expected value of the sample mass when the expected values $N'_i$ are proportional to the expected value of the sample mass. It can thus be seen that Gy's method to derive a model equation for the fundamental sampling error relies on the implicit assumption that first-order inclusion probabilities play a dominant role in explaining the sample-to-sample variations and that the higher second-order inclusion probabilities do not significantly influence the results, i.e. the second term, linear in $C_{ij}$, is negligible. The generalization therefore offers the potential of a more accurate description of the variance of the fundamental sampling error. The second term on the right-hand side can thus be considered to be a correction term, describing the effect of non-zero second-order inclusion probabilities. It is noted that this correction term may also depend on the expected value of the sample mass. Therefore, an easy determination of the correction term by considering a series of samples with large sample masses, such that the first term on the right-hand side would become negligible compared to the correction term, is not generally possible. The observation that the variance of the sampling error varies inversely proportional with the expected value of the sample mass should not be seen as an indication that the correction term (and hence the effect of non-zero second-order inclusion probabilities on the sampling variance) does not significantly influence the sampling variance.

Under certain assumptions about the sampling process, Lyman (1998) derived an expression for the sampling variance for sampling from a batch of particles in a totally segregated state and assumed that this value represents a maximum possible value for the sampling variance. It would be interesting to investigate whether (2) can give a theoretical basis for this assumption. Further discussion of this subject is however outside the scope of this article.

If the expected values $M'_s$, $N'_i$ and $c'_s$ are not known (as is generally the case in practice), the variance may be estimated by replacing these expected values in (2) by the corresponding values in a sample. The resulting variance estimator is then denoted as $Var(c_{sample})$ (instead of $V(c_{sample})$) and the corresponding sample values of $M'_s$, $N'_i$ and $c'_s$ are respectively denoted as $M_s$, $N_i$ and $c_s$. The thus obtained equation is:

$$\text{Var}(c_{sample}) = \frac{1}{M_s^2} \sum_i N_i m_i^2 (c_i - c_s)^2 - \frac{1}{M_s^2} \sum_i \sum_j C_{ij} N_i N_j m_i m_j (c_i - c_s)(c_j - c_s)$$

(Eq. 3)

where $M_s$, $N_i$ and $c_s$ represent respectively the mass of the sample, the number of particles belonging to the i-th class in the sample and the concentration in the sample. Another way of estimating the variance is based on the use of the Horvitz-Thompson estimator (see e.g. Särndal et al, 1992), which leads to the following variance estimator (Geelhoed, 2006), denoted as $V_{HT}(c_{sample})$:

$$V_{HT}(c_{sample}) = \frac{1}{M_s^2} \sum_i \frac{N_i m_i^2 c_i^2}{1 - C_{ii}} - \frac{1}{M_s^2} \sum_i \sum_j \frac{C_{ij} N_i N_j c_i c_j m_i m_j}{1 - C_{ij}}$$

(Eq. 4)

A derivation is presented in the Appendix. The equation is based on the additional assumption that the sampling process used to draw the sample leads to samples of a constant mass and that the first-order inclusion probability of the particles is given by the ratio of the sample mass and the mass of the batch from which the sample was drawn (i.e. sampling is correct). (4) is therefore not applicable when $C_{ij}=0$ for all possible values of i and j (Gy's model), because in that case the sample mass could not be constant, but would vary between zero and the mass of the batch from which the sample was drawn. If the sample values $M_s$, $N_i$ and $c_s$ are not known, these may be replaced by their expected values $M'_s$, $N'_i$ and $c'_s$ respectively, if available, in order to arrive at an estimated value for $V_{HT}$. It should be noted that the use of (4) can lead to different estimates for the variances than the use of (3), so future practical use of these equations needs to take into account the strengths and weaknesses of both options. A full discussion of this is however outside the scope of this paper. However, a brief and preliminary discussion will be given.

A general advantage of $V_{HT}(c_{sample})$ is that, because it is based on the Horvitz-Thompson estimator, which is unbiased, it is expected that the bias in $V_{HT}(c_{sample})$ is generally smaller than the bias in $\text{Var}(c_{sample})$. Another advantage of the use of $V_{HT}(c_{sample})$ is that, in the simple circumstance where there is only one particle class k consisting of particles with a non-zero value for the concentration $c_k$, while all other classes (i≠k) have $c_i=0$, the equation is simplified so that it does not explicitly depend on the masses of the other particles: $V_{HT}(c_{sample}) = (1/M_s)(c_s(1-N_k C_{kk})c_k m_k/(1-C_{kk}))$. If a reliable empirical estimate for the sampling variance, denoted here as $V_e$ is available, e.g. determined using the analysis of a series of samples of constant mass, this estimate can be substituted into the above equation, from which the parameter $C_{kk}$ can then be solved: $C_{kk}=(V_e-c_s c_k m_k/M_s)/(V_e-N_k c_s c_k m_k/M_s)$. Hence, $C_{kk}$ is zero when $V_e = c_s c_k m_k/M_s$, which will here be denoted as $V_{GY}$, because in Gy's sampling model $C_{kk}$ would be zero by definition. Table 1 shows the thus obtained estimate for $C_{kk}$ for several values of the ratio of $V_e$ and $V_{GY}$ and the number $N_k$ of particles with non-zero concentration in the sample (or its expected value $N'_k$).

Table 1 - Values of the estimated value of $C_{kk}$ for some values of the ratio of the empirical variance estimate $V_e$ and the variance estimate $V_{GY}$ and (the expected value of) the number of particles with non-zero concentration in the sample for the simple case of only one particle class (class k) with non-zero concentration.

| $N_k, N'_k$　$V_e/V_{GY} \rightarrow$ ↓ | 0.1 | 0.2 | 0.4 | 1 | 2 | 4 |
|---|---|---|---|---|---|---|
| 10 | $9.1 \times 10^{-2}$ | $8.1 \times 10^{-2}$ | $6.3 \times 10^{-2}$ | 0 | $-1.3 \times 10^{-1}$ | $-5.0 \times 10^{-1}$ |
| 100 | $9.0 \times 10^{-3}$ | $8.0 \times 10^{-3}$ | $6.0 \times 10^{-3}$ | 0 | $-1.0 \times 10^{-2}$ | $-3.1 \times 10^{-2}$ |
| 1000 | $9.0 \times 10^{-4}$ | $8.0 \times 10^{-4}$ | $6.0 \times 10^{-4}$ | 0 | $-1.0 \times 10^{-3}$ | $-3.0 \times 10^{-3}$ |
| 10000 | $9.0 \times 10^{-5}$ | $8.0 \times 10^{-5}$ | $6.0 \times 10^{-5}$ | 0 | $-1.0 \times 10^{-4}$ | $-3.0 \times 10^{-4}$ |

A general drawback of the use of $V_{HT}(c_{sample})$ is that it is strictly only applicable when the sample mass is constant. While it is technically possible to draw samples of approximately constant mass, remaining variations in sample mass could lead to a bias in the estimates calculated using $V_{HT}(c_{sample})$. Therefore, further study is required into practically acceptable values for the bias in variance estimates and acceptable levels of variation in sample mass for which $V_{HT}(c_{sample})$ can still be used without bias correction.

Three general advantages of the use of $Var(c_{sample})$ will be discussed. The first advantage of the use of $Var(c_{sample})$ is that it can easily be seen that the value of $Var(c_{sample})$ will be zero when all concentrations $c_i$ and $c_s$ are equal. This is desirable, because if all particles have the same concentration, the variance of the sample concentration will be zero. A second advantage is that the value of $Var(c_{sample})$ is not influenced by constant systematic errors in the determinations of the parameters $c_i$ and $c_s$. Finally, the third advantage is that $Var(c_{sample})$ depends in a linear way on the parameter $C_{ij}$, making it less sensitive to errors in the determination of $C_{ij}$ than $V_{HT}(c_{sample})$, for values of $C_{ij}$ close to one. A further discussion of the differences between the estimators expressed in (3) and (4) and the potential practical consequences is outside the scope of this article.

A prerequisite of application of the equations of the generalized theory is knowledge of the parameter for the dependent selection of particles, $C_{ij}$. It is therefore useful to identify the spatial distribution of particles in the batch before the sample is drawn as a dominant underlying source of dependent selections of particles (Geelhoed, 2005B). Two types of particles that tend to be more in the vicinity of each other than would be expected on the basis of completely independently randomly spatial distribution of particles will have an increased second-order inclusion probability and will therefore have a negative value of $C_{ij}$. On the other hand, two types of particles that tend to be further away from each other than would be expected on the basis of a completely random and independent spatial distribution of particles will have a decreased second-order inclusion probability and will therefore have a positive value of $C_{ij}$.

In view of the above remarks, factors contributing to a potential decrease of the value of $C_{ij}$ are identified as: clustering of particles caused by attractive inter-particle forces, such as attractive electrostatic forces or moisture that makes particles wet and stick to each other. Factors leading to a potential increase in the value of $C_{ij}$ are identified as: density differences combined with the influence of gravity during transport or repulsive inter-particle forces. With all of the effects mentioned, the size and shape of the particles also influence the magnitude of the effect. Size is also in itself a factor leading to a potential increase of the value of $C_{ij}$. This can be demonstrated using a simple example of two spherical particles with diameter $D_1$ and $D_2$. The distance between (the centres of masses of) the particles cannot be smaller than $D_1/2 + D_2/2$. Hence, larger particles will on average be further away from each other than smaller particles, leading to a potential increase of $C_{ij}$.

## RESULTS

A generalization of the theory of Gy was reviewed, which can potentially lead to more accurate estimates for the variance of the fundamental sampling error. Because a prerequisite of application of the equations of the generalized theory is knowledge of the parameter for the dependent selection of particles, an important result is the identification of the spatial distribution of particles in the batch before the sample is drawn as a dominant underlying source of dependent selections of particles. This then results in the identification of clustering of particles, differences in densities or sizes of the particles and repulsive inter-particle forces as factors influencing the magnitude of the parameter for the dependent selection of particles.

## DISCUSSION

Because the generalization has the potential of providing more accurate predictions for the variance of the fundamental error, possible approaches to evaluate the essential new input parameter for the generalized theory, the parameter for the dependent selection of particles, are discussed below.

A first approach would be to find a model equation for predicting the value of $C_{ij}$ (and ultimately also of the variance) based on the physical properties of the particles such as size, density, shape and electrostatic properties. This is a difficult and complex task, which, if possible, may be useful for certain applications, but will not be further discussed here. A second approach would be the application of image analysis of samples, which allows observing directly the spatial distribution of particles. Although it is generally easy to obtain an image, there are many potential ways of extracting information about the value of the parameter $C_{ij}$ from the image. Below, one possible way will be discussed.

A possible way of evaluating the parameter for the dependent selection of particles would be to use "line-intercept sampling" (also known as "line-transect sampling") of the image, a method of sampling particles in a region whereby, roughly, a particle is sampled if a chosen line segment, called a "transect", intersects the particle (Kaiser, 1983). This produces a one-dimensional chain of particles from which information about the spatial distribution, specifically $C_{ij}$ can be derived. This is done by first counting the numbers of transitions between the different particle types. The number of transitions going from a particle of type i to a particle of type j is denoted as $N_{ij}$. Intuitively it is clear that there will be a relationship between $C_{ij}$ and $N_{ij}$: a large negative value of $C_{ij}$ will lead to a high value of $N_{ij}$, because particle types that have a high second-order inclusion probability tend to have more transitions between each other. Markov Chain modelling (see e.g. Freedman, 1971) could be applied to quantify this relationship. However, this will not be further developed here.

A potential problem with the above-described method is that the line-intercept sample could be biased towards larger particles, because of the increased probability of intersecting a larger particle. Therefore, further study is required into the effect of this potential bias on the estimates for $C_{ij}$ and its effect on the variance estimate. Also further study into ways of overcoming this bias is required for cases where the effect of this bias on the variance estimate is non-negligible.

Some general observations concerning the second approach can be made based on the fact that this approach relies on an image of the sample. Because two-dimensional images can be obtained using a variety of routine techniques ranging from digital photography to Scanning Electron Microscopy, further study into the feasibility of the second approach is required for a range of materials and techniques to obtain two-dimensional images. A common problem with these 2D techniques, however, is that a surface sample may be unrepresentative for the whole sample. This problem can be overcome by using a 3D imaging technique (if available of course!) or using multiple two-dimensional cross sections of the same sample. However, for fast and cheap results, using a digital image of the top surface would be ideal. Therefore, further research is also proposed into this application.

A final point of discussion concerning the generalized theory as a whole is that although it has the potential of providing more accurate variance estimates than the theory of Gy, practically relevant particle systems must be identified for which it can be demonstrated that this is a significant improvement. Further experimental work is therefore required before the new approach can become a routinely applied and generally accepted method.

## CONCLUSIONS

At the moment it is too early for a routine application of the generalized theory. However, the generalization has the potential of providing more accurate variance estimates than is now possible in the theory of Gy. In view of this, two approaches for evaluating an essential input parameter of the generalized model are discussed: a modelling approach based on the particle properties and an approach based on image analysis of an actual sample. Only one method, a method based on line-intercept sampling and Markov Chain modelling, in the second approach is briefly discussed in this article. There are many other potential methods of determining the parameter for the dependent selection of particles. Because the generalized theory can potentially lead to more accurate estimates for the variance of the fundamental sampling error, further research into the development of new methods in both approaches is proposed.

## APPENDIX

A $\pi$-expanded estimator (see e.g. Särndal et al, 1992) for the concentration in the batch is given by:

$$<c_{batch}>_\pi = \sum_i N_i m_i c_i / M_{batch} \pi_i \qquad (Eq.\ 5)$$

where $<c_{batch}>_\pi$ is the $\pi$-expanded estimator for the concentration in the batch, $N_i$ is the number of particles in the sample belonging to the i-th particle class, $m_i$ and $c_i$ are respectively the mass of and the concentration in a particle belonging to the i-th class, $M_{batch}$ is the mass of the batch and $\pi_i$ is the first-order inclusion probability of a particle belonging to the i-th class. A derivation of the above equation can be found in Geelhoed (2004). If the sample mass is constant and the first-order inclusion probability is equal to the ratio of the sample mass ($M_s$) and the batch mass ($M_{batch}$), the $\pi$-expanded estimator becomes equal to the sample concentration, $c_s$. Under these assumptions (and conditions (i), (iii) and (iv) stated in the main text of this article), Geelhoed (2004) derived the following equation for the variance of the sample concentration, based on the general Horvitz-Thompson estimator for the variance of the $\pi$-expanded estimator:

$$V_{HT}(c_{sample}) = \sum_i \sum_j N_i N_j (\frac{1}{\pi_i \pi_j} - \frac{1}{\pi_{ij}}) \frac{m_i m_j c_i c_j}{M_{batch}^2} + \sum_i N_i (\frac{1}{\pi_{ii}} - \frac{1}{\pi_i}) \frac{m_i^2 c_i^2}{M_{batch}^2} \qquad (Eq.\ 6)$$

in which $\pi_{ij}$ is the second-order inclusion probability of a particle pair in which the first-particle belongs to the i-th class and the second to the j-th class. Substitution of (1) for the second-order inclusion probability and $\pi_i = M_s/M_{batch}$, results in:

$$V_{HT}(c_{sample}) = \sum_i \sum_j N_i N_j (1 - \frac{1}{1-C_{ij}}) \frac{m_i m_j c_i c_j}{M_s^2} + \sum_i (\frac{1}{1-C_{ii}} - \frac{M_s}{M_{batch}}) N_i \frac{m_i^2 c_i^2}{M_s^2} \quad \text{(Eq. 7)}$$

Assuming that the batch from which the sample was drawn is much larger than the sample, i.e. $M_{batch} \gg M_{sample}$ so that $1/(1-C_{ii}) - M_s/M_{batch} \approx 1/(1-C_{ii})$, the above result can be rearranged to yield (4) in the main text.